\documentclass[aps,prd,amsmath,amssymb,showpacs,10pt,superscriptaddress]{revtex4-2}
\usepackage{graphicx}
\usepackage{color}
\usepackage{slashed}
\usepackage{txfonts}
\usepackage{multirow}
\usepackage{subfigure}
\usepackage{enumerate}
\usepackage{physics}
\usepackage[font=footnotesize]{caption}
\usepackage[breaklinks,colorlinks,urlcolor=blue,linkcolor=red,anchorcolor=magenta,citecolor=blue]{hyperref}
\usepackage{cleveref}
\crefname{table}{Table}{Tables}
\crefname{equation}{Eq.}{Eqs.}
\crefname{figure}{Fig.}{Figs.}
\crefname{section}{Sec.}{Secs.}
\allowdisplaybreaks
\newcommand{\ihep}{\affiliation{Institute of High Energy Physics, Chinese Academy of Sciences, Beijing 100049, People's Republic of China}}
\newcommand{\ucas}{\affiliation{University of Chinese Academy of Sciences, Beijing 100049, People's Republic of China}}
\begin{document}
\title{Identification of $D_{s1}(2933)$ as the $2P_{1}$ state in the quark model}
\author{Jun Wang}\email{junwang@ihep.ac.cn}
\ihep\ucas
\author{Qiang Zhao}\email{zhaoq@ihep.ac.cn}
\ihep\ucas
\begin{abstract}
  We investigate the internal structure of the newly observed charm-strange meson $D_{s1}(2933)^{+}$, which was reported by the LHCb Collaboration in  $B^{0}\to D^{+}D^{-}K^{+}\pi^{-}$, in the quark model by treating it as the $2P_{1}$ state of the $c\bar{s}$ system. The coupled-channel corrections to the ${}^{1}P_1$--${}^{3}P_1$ mass matrix are evaluated through the two-point self-energy loops involving the $DK^{*}$, $D^{*}K$, $D^{*}K^{*}$, $D_{s}\phi$, $D_{s}^{*}\phi$ and $D_{s}^{*}\eta$ intermediate channels. Requiring the corrected eigenvalue to reproduce the measured mass yields a cutoff scale of about $0.82~\mathrm{GeV}$, a $2P_{1}^{\prime}$ partner state at $3012~\mathrm{MeV}$, and a mixing angle of $-12.9^{\circ}$ that deviates significantly from the heavy-quark ideal-mixing angle. Moreover, the predicted three-body branching ratio fraction $R_{\text{th}}=1.98$ for $D_{s1}(2933)\to D^{+}K^{+}\pi^{-}$ via the $DK^{*}$ and $D_{2}^{*}K$ intermediate states is consistent with the experimental value $R_{\text{exp}}=1.27^{+1.29}_{-0.95}$ within uncertainties. These results support the assignment of $D_{s1}(2933)$ as the $2P_{1}$ excited state of the $c\bar{s}$ system with final state interactions.
\end{abstract}

\maketitle

\section{Introduction}\label{sec:introduction}
Since 2003, the successive discoveries of positive-parity charm-strange mesons such as $D_{s0}^{*}(2317)$~\cite{BaBar:2003oey} and $D_{s1}(2460)$~\cite{CLEO:2003ggt}, as well as the subsequent discoveries of higher excited states like $D_{sJ}(2700)$~\cite{Belle:2007hht} and $D_{sJ}(2860)$~\cite{LHCb:2014ott}, have made charm-strange meson spectroscopy one of the core issues in hadron physics~\cite{Godfrey:2015dva,Guo:2017jvc,ParticleDataGroup:2024cfk}. The masses, widths, and decay properties of these states deviate significantly from the predictions of the conventional $c\bar{s}$ quark model~\cite{Godfrey:1985xj,Godfrey:1986wj,Godfrey:2015dva}, stimulating extensive discussions on non-trivial structures such as coupled-channel effects, hadronic molecules, and tetraquark states~\cite{Guo:2011dd,Molina:2010tx,Hao:2022vwt,Chen:2016qju,Chen:2022asf,Guo:2017jvc,Zhu:2007wz,Liu:2019zoy,Zhu:2004xa}. Therefore, new experimental signals can often provide crucial clues for clarifying the internal structures of these excited states.

Recently, through a full phase-space amplitude analysis of the $B^{0}\to D^{+}D^{-}K^{+}\pi^{-}$ decay process, the LHCb Collaboration simultaneously observed a new excited charm-strange meson $D_{s1}(2933)^{+}$ in two different quasi-two-body decay channels, $D^{+}K^{*0}$ and $K^{+}D_{2}^{*0}$, via particle reconstruction. Its mass is about $2933~\mathrm{MeV}$, and its spin-parity quantum numbers are $J^{P}=1^{+}$~\cite{LHCb:2026sup}. On the one hand, its mass is about $80$--$100~\mathrm{MeV}$ lower than the lowest $2P_{1}$ state prediction ($\sim 3.02~\mathrm{GeV}$) given by the Godfrey-Isgur (GI) quark model~\cite{Godfrey:2015dva,Godfrey:1985xj,Godfrey:1986wj}, a discrepancy similar to those observed for $D_{s0}^{*}(2317)$ and $D_{s1}(2460)$~\cite{Guo:2017jvc,Hao:2022vwt,vanBeveren:2003kd}. On the other hand, $D_{s1}(2933)$ lies near the thresholds of the $D^{*}K^{*}$ and $D^{(*)}_{s}\phi$ channels, which naturally suggests that coupled-channel effects or near-threshold molecular components may play an important role~\cite{Guo:2017jvc,Barnes:2003dj}.

Regarding the internal structure of $D_{s1}(2933)$, the conventional $c\bar{s}$ quark model picture and the hadronic molecular picture are the two main candidate scenarios. In the conventional quark model scenario, since the spin-parity of $D_{s1}(2933)$ is $1^+$, and its energy is significantly higher than the $1P$ multiplet, it is natural to assign it as a radially excited $2P_{1}$ state~\cite{Godfrey:2015dva,Godfrey:2014fga}. However, to explain the discrepancy with the quark model eigenstates, which are also called ``bare states'', one must systematically consider the unquenched corrections induced by intermediate meson loops such as $DK^{*}$, $D^{*}K$, $D^{*}K^{*}$, $D_{s}\phi$, $D_{s}^{*}\phi$, and $D_{s}^{*}\eta$~\cite{Hao:2022vwt,Barnes:2007xu,Liu:2020ruo,Boglione:2002vv,Rupp:2006sb,Wang:2025wpc}. Such coupled-channel effects can not only lead to a significant downward shift of the bare-state mass but also change the mixing angle between $^{1}P_{1}$ and $^{3}P_{1}$, thereby affecting the partial widths of $D_{s1}(2933)$ decays into channels such as $DK^{*}$, $D^{*}K$, and $D_{s}^{*}\eta$~\cite{Godfrey:2015dva,Liu:2020ruo}. Similar phenomena also appear in the interpretation of $D_{s1}(2460)$ and $D_{s1}(2536)$ as physical mixtures of the ${}^{1}P_1$ and ${}^{3}P_1$ states~\cite{Wu:2011yb}. On the other hand, the proximity of $D_{s1}(2460)$ to the $D^{*}K$ threshold has motivated its interpretation as a $D^{*}K$ hadronic molecule, while coupled-channel effects can play an important role in both $D_{s1}(2460)$ and $D_{s1}(2536)$~\cite{Wu:2011yb,Molina:2010tx}. In the literature, $D_{s1}(2460)$ has also been considered as a tetraquark candidate~\cite{Maiani:2004vq}. To understand whether $D_{s1}(2933)$ is an ordinary $2P_{1}$ state or contains a significant hadronic molecular component, it is necessary to simultaneously consider its mass and decay branching ratios in a unified framework.

In this work, we treat $D_{s1}(2933)$ as one of the $2P_{1}$ excited states of the $c\bar{s}$ system, and systematically calculate the effects of coupled-channel corrections on the $2\,{}^{1}P_{1}-2\,{}^{3}P_{1}$ mass matrix. Based on the GI quark model framework~\cite{Godfrey:2015dva,Godfrey:1985xj,Godfrey:1986wj} the mixing angle can be extracted through the spin-orbital quark-quark interaction, and the strong coupling constants can be extracted in the framework of the $^{3}P_{0}$ model for the meson decays~\cite{Micu:1968mk,LeYaouanc:1972vsx,LeYaouanc:1973ldf,LeYaouanc:1988fx,Ackleh:1996yt,Blundell:1996as,Segovia:2012cd}. It should be mentioned that the spin-orbital quark-quark interaction in the GI is suppressed by $1/m_q^2$ which leads to an ideal mixing pattern for most of the quark model calculations for the axial vector meson mixings. Such a pattern can be violated if nearby strong coupled-channel effects are present as mentioned earlier. Thus, we consider contributions from the $DK^{*}$, $D^{*}K$, $D^{*}K^{*}$, $D_{s}\phi$, $D_{s}^{*}\phi$ and $D_{s}^{*}\eta$ intermediate channels via two-point self-energy diagrams~\cite{Barnes:2007xu,Hao:2022vwt} as an additional mechanism for the $2\,{}^{1}P_{1}-2\,{}^{3}P_{1}$ mixing. By requiring the lower corrected eigenvalue to reproduce the experimentally measured physical-state mass, we can extract the mass matrix and the corresponding mixing angle.
Based on the obtained mixing angle and physical-state wave functions, we can calculate the total width of $D_{s1}(2933)$ by evaluating the partial widths of the dominant decay channels $DK^{*}$, $D^{*}K$, $D^{*}K^{*}$ and $D_{s}^{*}\eta$, and then compare it with the experimental width.
Furthermore, we can evaluate the relative three-body decay branching ratio $R_{\text{th}}$ of $D_{s1}(2933)$ to $D^{+}K^{+}\pi^{-}$ via the $DK^{*}$ and $D_{2}^{*}K$ intermediate states using the $^{3}P_{0}$ model, and compare it with the experimental ratio $R_{\text{exp}}$ extracted by LHCb in the $B^{0}\to D^{+}D^{-}K^{+}\pi^{-}$ process. This provides a complementary test of the assignment of $D_{s1}(2933)$ as the $2P_{1}$ state.

The remainder of this paper is organized as follows. We first introduce our formalism in \cref{sec:formalism}. The numerical results and discussion are presented in \cref{sec:results}. A brief summary and conclusion are given in \cref{sec:summary}.

\section{Formalism}\label{sec:formalism}
\subsection{Mass shift in the quark model}
The $2P_{1}^{(\prime)}$ states predicted by the GI model have masses of $3.018~(3.038)~\mathrm{GeV}$~\cite{Godfrey:2015dva,Godfrey:1985xj,Godfrey:1986wj}, which are higher than the experimentally observed mass of $D_{s1}(2933)$. Given the relatively low mass of $D_{s1}(2933)$, we identify it as a candidate for the lower-mass $2P_{1}$ state. We consider the coupled-channel corrections to the bare-state masses of the ${}^{1}P_1$ and ${}^{3}P_1$ states predicted by the quark model, and solve for the eigenvalues of the resulting mass matrix to obtain the physical-state masses. The bare-state mass matrix is obtained from the physical-state masses and mixing angle predicted by the GI model via a similarity transformation,
\begin{equation}
  M^{0}=\mqty(\mel*{{}^{1}P_1}{\mathcal{H}^{0}}{{}^{1}P_1}&\mel*{{}^{1}P_1}{\mathcal{H}^{1}}{{}^{3}P_1}\\\mel*{{}^{3}P_1}{\mathcal{H}^{1}}{{}^{1}P_1}&\mel*{{}^{3}P_1}{\mathcal{H}^{0}}{{}^{3}P_1})=\mathcal{P} M_{\text{diag}}\mathcal{P}^{-1}\,,
\end{equation}
where $\mathcal{H}^{0}$ is the Hamiltonian in the GI model and $\mathcal{H}^{1}$ is the spin-orbit coupling term, $\mathcal{P}$ is the mixing matrix and $M_{\text{diag}}$ is the diagonal matrix of physical-state masses predicted by the quark model. The mixing matrix and mixing angle are defined as follows:
\begin{equation}
  \begin{aligned}
    \ket{P}= &\cos \theta_{P}\ket*{{}^{1}P_1}+ \sin \theta_{P}\ket*{{}^{3}P_1} \\
    \ket*{P^{\prime}}= &-\sin \theta_{P}\ket*{{}^{1}P_1}+ \cos \theta_{P}\ket*{{}^{3}P_1}
  \end{aligned},\quad
  \mathcal{P}=\mqty(\cos \theta_{P}&-\sin \theta_{P}\\\sin \theta_{P}&\cos \theta_{P})\,.
\end{equation}

\begin{figure}[!h]
  \centering
  \subfigure[]{\includegraphics[width=0.45\textwidth]{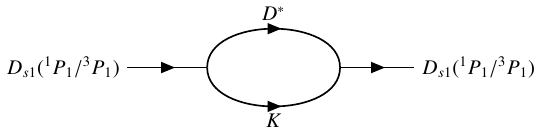}}
  \subfigure[]{\includegraphics[width=0.45\textwidth]{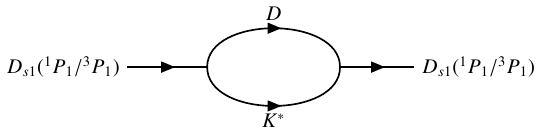}}\\
  \subfigure[]{\includegraphics[width=0.45\textwidth]{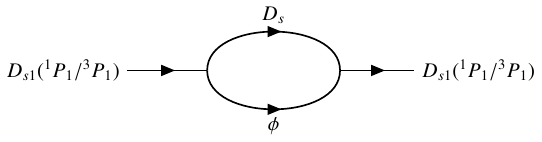}}
  \subfigure[]{\includegraphics[width=0.45\textwidth]{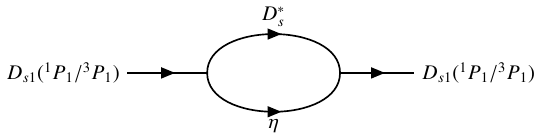}}\\
  \subfigure[]{\includegraphics[width=0.45\textwidth]{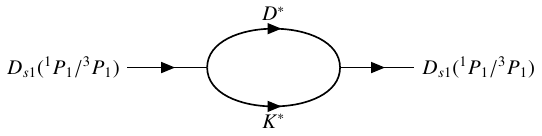}}
  \subfigure[]{\includegraphics[width=0.45\textwidth]{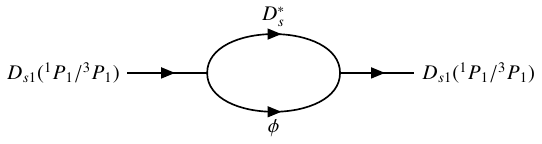}}
  \caption{Schematic diagrams of the two-point loop contributions to the mass matrix elements. (a)--(f) represent different intermediate meson loops.
  }\label{fig1}
\end{figure}

In the GI model, the ${}^{1}P_1$-${}^{3}P_1$ mixing is induced by the spin-orbit coupling term, while we calculate the physical-state mass and mixing angle of $D_{s1}(2933)$ by including coupled-channel corrections.
The coupled-channel mass corrections to the bare states are obtained by evaluating the two-point loop diagrams, as shown in \cref{fig1}.

The corresponding effective Lagrangians are
\begin{equation}
  \begin{aligned}
    \mathcal{L}_{D_{s1}D^{*}K}=&g_{D_{s1}D^{*}K}D_{s1}^{\mu}D^{*}_{\mu}K,\quad \mathcal{L}_{D_{s1}DK^{*}}=g_{D_{s1}DK^{*}}D_{s1}^{\mu}K_{\mu}^{*}D,\quad
    \mathcal{L}_{D_{s1}D_{s}\phi}=g_{D_{s1}D_{s}\phi}D_{s1}^{\mu}D_{s}\phi_{\mu},\quad \mathcal{L}_{D_{s1}D_{s}^{*}\eta}=g_{D_{s1}D_{s}^{*}\eta}D_{s1}^{\mu}D^{*}_{s\mu}\eta,\\
    \mathcal{L}_{D_{s1}D^{*}K^{*}}=&ig_{D_{s1}D^{*}K^{*}}\epsilon_{\mu\nu\alpha\beta}\partial^{\mu}D_{s1}^{\nu}D^{*\alpha}K^{*\beta},\quad \mathcal{L}_{D_{s1}D_{s}^{*}\phi}=ig_{D_{s1}D_{s}^{*}\phi}\epsilon_{\mu\nu\alpha\beta}\partial^{\mu}D_{s1}^{\nu}D_{s}^{*\alpha}\phi^{\beta}\,.\label{eq:el1}
  \end{aligned}
\end{equation}
We only consider the $S$-wave contributions; the corresponding coupling constants are determined from the bare-state masses using the $^{3}P_{0}$ model~\cite{LeYaouanc:1973ldf,LeYaouanc:1972vsx}.

Based on the effective Lagrangians above, we can write down the two-point loop amplitudes for $D_{s1}$.
For the process $D_{s1}(p) \to V(q) + P(p-q) \to D_{s1}(p)$, where $V$ denotes an intermediate vector meson and $P$ an intermediate pseudoscalar meson, the two-point self-energy correction $\Sigma_{VP}^{\mu\nu}(p)$ takes the unified form:
\begin{equation}
  \Sigma_{VP}^{\mu\nu}(p) = i g_{VP}g_{VP}^{\prime} \int \frac{d^4 q}{(2\pi)^4} \frac{-g^{\mu\nu} + \frac{q^\mu q^\nu}{m_{V}^{2}}}{(q^2 - m_V^2 + i\epsilon)[(p-q)^2 - m_P^2 + i\epsilon]} \mathcal{F}(q^2)\,,
\end{equation}
where $\mathcal{F}(q^2)$ is a form factor introduced to regularize the ultraviolet divergence,
\begin{equation}
  \mathcal{F}(q^{2})=\frac{\Lambda^{2}-m_{V}^{2}}{\Lambda^{2}-q^{2}}\frac{\Lambda^{2}-m_{P}^{2}}{\Lambda^{2}-(q-p)^{2}},\quad \Lambda=m_{V}(m_{P})+\alpha \Lambda_{\mathrm{QCD}},\quad \Lambda_{\mathrm{QCD}}=220~\mathrm{MeV}.
\end{equation}
Similarly, for the process $D_{s1}(p)\to V_{1}(q)+V_{2}(p-q)\to D_{s1}(p)$, the corresponding self-energy correction is
\begin{equation}
  \Sigma_{VV}^{\mu\nu}(p)=i g_{V_1V_2}g_{V_1V_2}'g^{\mu \alpha}g^{\delta \nu}\int \frac{d^{4}q}{(2\pi)^{4}}\frac{\varepsilon_{\alpha \beta \rho\sigma}\varepsilon_{\delta \zeta \tau \kappa}p^{\beta}p^{\zeta}\qty(-g^{\rho \tau}+\frac{q^{\rho}q^{\tau}}{m_{V_1}^{2}})\qty(- g^{\sigma \kappa}+\frac{(p-q)^{\sigma}(p-q)^{\kappa}}{m_{V_2}^{2}})}{(q^{2}-m_{V_{1}}^{2}+i\epsilon)((p-q)^{2}-m_{V_{2}}^{2}+i \epsilon)}\mathcal{F}(q^{2})\,.
\end{equation}

For different $VP$ intermediate-state processes ($D^{*}K$, $DK^{*}$, $D_{s}\phi$ and $D_{s}^{*}\eta$), or $VV$ intermediate-state processes ($D^{*}K^{*}$ and $D_{s}^{*}\phi$), one only needs to replace the coupling constants $g_{VP}$ ($g_{V_1V_2}$) and the meson masses $m_V, m_P$ ($m_{V_1}, m_{V_2}$) in the above expressions with the specific values for the corresponding channels. $D_{s1}$ is a spin-1 particle; its self-energy admits a tensor decomposition into transverse and longitudinal components:
\begin{equation}
  \Sigma^{\mu\nu}(p) = \left( -g^{\mu\nu} + \frac{p^\mu p^\nu}{p^2} \right) \Sigma_T(p^2) + \frac{p^\mu p^\nu}{p^2} \Sigma_L(p^2)\,.
\end{equation}
The physically relevant mass correction comes from the transverse part $\Sigma_T(p^2)$, which is projected out via
\begin{equation}
  \Sigma_T(p^2) = -\frac{1}{3} \left( g_{\mu\nu} - \frac{p_\mu p_\nu}{p^2} \right) \Sigma^{\mu\nu}(p)\,.
\end{equation}
The corresponding mass matrix element induced by coupled-channel corrections reads~\cite{Barnes:2007xu}
\begin{equation}
  M_{ij}(m)=M^{0}_{ij}\delta_{ij}+\frac{\operatorname{Re}\Sigma_{ij}^{T}(m^{2})}{2m}\,,
\end{equation}
here, $\delta_{ij}$ indicates that the mixing induced by the off-diagonal terms arises only from coupled-channel corrections, while the diagonal elements receive contributions from both the bare-state masses and the coupled-channel corrections.

We adopt a fixed-reference-mass prescription, evaluating the self-energy corrections at $m=2933~\mathrm{MeV}$. At this reference mass, the cutoff parameter $\alpha$ is determined by imposing the secular condition
\begin{equation}
  \det(M_{ij}(m)-m\delta_{ij})=0\,. \label{eq:detm}
\end{equation}
For our primary assignment, we select the value of $\alpha$ for which the lower eigenvalue of the resulting effective mass matrix equals the reference mass. Once $\alpha$ is fixed, the partner-state mass and mixing angle are obtained by diagonalizing this same matrix, without re-evaluating the self-energy at the partner-state mass. The alternative assignment of $D_{s1}(2933)$ as the higher-mass $2P_{1}^{\prime}$ state is examined analogously by requiring the higher eigenvalue to equal the same reference mass while keeping all other model inputs unchanged.

\subsection{$^{3}P_{0}$ model}
The transition operator in the $^{3}P_{0}$ model is~\cite{Ackleh:1996yt,Blundell:1996as,LeYaouanc:1972vsx,Micu:1968mk,LeYaouanc:1988fx}
\begin{equation}
  T = -3 \gamma \sum_{m} \innerproduct{1,m;1,-m}{0,0} \int d^{3}\mathbf{p}_{3} d^{3}\mathbf{p}_{4} \, \delta^{3}(\mathbf{p}_{3}+\mathbf{p}_{4}) \, \mathcal{Y}_{1}^{m}\qty(\frac{\mathbf{p}_{3}-\mathbf{p}_{4}}{2}) \chi^{34}_{1,-m} \phi^{34}_{0} \omega^{34}_{0} b^{\dagger}_{3i}(\mathbf{p}_{3}) d^{\dagger}_{4j}(\mathbf{p}_{4})\,,
\end{equation}
where $i$ and $j$ are color indices, $\chi_{1,-m}^{34}$ is the spin wave function, $\phi_{0}^{34}$ is the flavor-singlet wave function, $\omega^{34}_{0} = \delta_{ij}/\sqrt{3}$ is the color-singlet wave function, and $\mathcal{Y}^{m}_{1}(\mathbf{p}) = |\mathbf{p}| Y^{m}_{1}(\theta,\phi)$. The parameter $\gamma$ is the $^{3}P_{0}$ pair-creation coupling constant.
The meson wave function is~\cite{Hayne:1981zy}
\begin{equation}
  \begin{aligned}
    \ket{M(\mathbf{P},J,J_{z})}=&\sum_{S_{z},L_{z},c_{i}}\innerproduct{L,L_{z};S,S_{z}}{J,J_{z}}\int d^{3}\mathbf{p}_{1} d^{3}\mathbf{p}_{2}\,\delta^{3}(\mathbf{p}_{1}+\mathbf{p}_{2}-\mathbf{P})\,\psi_{N,L,L_{z}}(\mathbf{p}_{1},\mathbf{p}_{2}) \\
    &\times \frac{\delta_{c_1c_2}}{\sqrt{3}}\phi_{f_1,f_2}\chi_{s_1,s_2}^{S,S_{z}}\,b^{\dagger}_{c_1,f_1,s_1,\mathbf{p}_{1}}\,d^{\dagger}_{c_2,f_2,s_2,\mathbf{p}_{2}}\ket{0}\,,
  \end{aligned}\label{eq:mesonwf}
\end{equation}
where $c_j$, $s_j$, and $f_j$ ($j=1,2$) are the color, spin, and flavor quantum-number indices of the quarks, respectively. $b^{\dagger}$ and $d^{\dagger}$ are the quark and antiquark creation operators, and $\psi_{N,L,L_{z}}$ is the spatial wave function in the harmonic-oscillator basis.
In the $^{3}P_{0}$ model, the transition amplitude for $A\to BC$ is
\begin{equation}
  \mathcal{M}_{q} = \mel{BC}{T}{A}\,.\label{eq:ampa}
\end{equation}
The amplitude obtained from the effective Lagrangian approach is denoted $\mathcal{M}_{h}$. The two amplitudes are related by $\mathcal{M}_{h} = 8\pi^{3/2} \sqrt{m_{A}m_{B} m_{C}}\,\mathcal{M}_{q}$, which allows the coupling constants in the effective Lagrangian to be determined. By substituting the wave functions from \cref{eq:mesonwf} into \cref{eq:ampa} and matching the resulting amplitude to the amplitude obtained from the effective Lagrangians in \cref{eq:el1}, the relevant vertex coupling constants can be extracted.
\subsection{Partial decay widths}
Once the physical-state masses and mixing angle are determined, we calculate the total width of $D_{s1}(2933)$. The dominant decay channels of $D_{s1}(2933)$ are $DK^{*}$, $D^{*}K$, $D^{*}K^{*}$ and $D_{s}^{*}\eta$. Using the $^{3}P_{0}$ model together with the extracted mixing angle, we compute the partial widths of these four channels to obtain the total width.

Furthermore, $D_{s1}(2933)$ was experimentally observed through the processes $B^{0}\to D^{-} D_{s1}(2933)\to D^{+}K^{*0}(\to K^{+}\pi^{-})$ and $B^{0}\to D^{-} D_{s1}(2933)\to K^{+}D_{2}^{*0}(\to D^{+}\pi^{-})$. The fit fractions of these two processes in $B^{0}\to D^{+}D^{-}K^{+}\pi^{-}$ were extracted experimentally, from which one can obtain the ratio
\begin{equation}
  R_{\text{exp}}= \frac{\mathrm{BR}(B^{0}\to D^{-} D_{s1}(2933)\to D^{+}K^{*0}(\to K^{+}\pi^{-}))}{\mathrm{BR}(B^{0}\to D^{-} D_{s1}(2933)\to K^{+}D_{2}^{*0}(\to D^{+}\pi^{-}))}\,.
\end{equation}
Since the production process is identical in both cases, $R_{\text{exp}}$ equals the relative three-body branching ratio of $D_{s1}(2933)$ decaying to $D^{+}K^{+}\pi^{-}$ via two different intermediate states. This ratio can be computed theoretically as
\begin{equation}
  R_{\text{th}}=\frac{\Gamma(D_{s1}(2933)\to D^{+}K^{*0}(\to K^{+}\pi^{-}))}{\Gamma(D_{s1}(2933)\to K^{+}D_{2}^{*0}(\to D^{+}\pi^{-}))}\,.
\end{equation}
A comparison of $R_{\text{th}}$ with $R_{\text{exp}}$ provides a test of our structural assignment for $D_{s1}(2933)$.
The relevant effective Lagrangians for computing $R_{\text{th}}$ are
\begin{equation}
  \begin{aligned}
    \mathcal{L}_{D_{s1}DK^{*}}=&g_{D_{s1}DK^{*}}D_{s1}^{\mu}K_{\mu}^{*}D, \quad \mathcal{L}_{K^{*}K\pi}=i g_{K^{*}K\pi}K^{*}_{\mu}(\partial^{\mu}K\pi-K\partial^{\mu}\pi),\\ \mathcal{L}_{D_{s1}D_{2}^{*}K}=&ig_{D_{s1}D_{2}^{*}K}D_{s1}^{\mu}D_{2\mu\nu}^{*}\partial^{\nu}K,\quad \mathcal{L}_{D^{*}_{2}D\pi}=g_{D_{2}^{*}D\pi}D^{*}_{2\mu\nu}\partial^{\mu}D\partial^{\nu}\pi\,, \label{eq:efl}
  \end{aligned}
\end{equation}
Here, the couplings $g_{D_{s1}DK^{*}}$ and $g_{D_{s1}D_{2}^{*}K}$ are determined from the $^{3}P_{0}$ model, whereas $g_{K^{*}K\pi}$ and $g_{D_{2}^{*}D\pi}$ are extracted from experimental decay information, as detailed in \cref{sec:results}. Since $D_{s1}(2933)$ lies below the $D_{2}^{*}K$ mass threshold, we adopt the on-shell approximation when solving for the coupling constants with the $^{3}P_{0}$ model.
\section{Numerical Results and Discussion}\label{sec:results}
We first employ the $^{3}P_{0}$ model to obtain the relevant coupling constants. The $^{3}P_{0}$ model coupling constant is taken as $\gamma=0.4$~\cite{Godfrey:2015dva}. The harmonic oscillator parameters and constituent quark masses are given in \cref{tab:qpcconstant}~\cite{Godfrey:2015dva,LeYaouanc:1972vsx,LeYaouanc:1973ldf,Segovia:2012cd}. The calculated coupling constants are summarized in \cref{tab:qpc1}.
It is worth noting that, in the ${}^{3}P_{0}$ model, the $S$-wave amplitude for the ${}^{3}P_{1}\to VV$ transition is proportional to the Wigner $9j$ symbol~\cite{Barnes:2002mu,Blundell:1996as,LeYaouanc:1972vsx,Ackleh:1996yt}
\begin{equation}
  \begin{Bmatrix}
    1/2& 1/2& 1\\
    1/2 & 1/2 & 1\\
    1 & 1 & 1
  \end{Bmatrix}=0\,.
\end{equation}
This symbol vanishes identically as a consequence of the spin-recoupling selection rule. Therefore, within an $S$-wave treatment, the $VV$ intermediate states do not contribute to the mass shift of the ${}^{3}P_{1}$ state. They can, however, contribute to the ${}^{1}P_{1}$ diagonal element, while the off-diagonal mass correction is generated solely by the $VP$ channels. Thus, the $VV$ channels can provide important diagonal mass shifts without inducing ${}^{1}P_{1}$--${}^{3}P_{1}$ mixing.

\begin{table}[!h]
  \centering\caption{Harmonic oscillator parameters and constituent quark masses used in the $^{3}P_{0}$ model.}
  \begin{ruledtabular}
    \begin{tabular}{cccc}
      HO strength & Values ($\mathrm{GeV}$) & Quark mass & Values ($\mathrm{GeV}$)\\\hline
      $\beta_{D_{s}(2\,{}^{1}P_1)}(\beta_{D_{s}(2\,{}^{3}P_1)})$&$0.433~(0.434)$&$m_{c}$&$1.628$\\
      $\beta_{D}(\beta_{D^{*}})$&$0.601~(0.516)$&$m_{s}$&$0.419$\\
      $\beta_{D_{s}}(\beta_{D_{s}^{*}})$&$0.651~(0.562)$&$m_{u/d}$&$0.22$\\
      $\beta_{D_{2}^{*}}$&$0.437$&
    \end{tabular}\label{tab:qpcconstant}
  \end{ruledtabular}
\end{table}

\begin{table}[!h]
  \centering\caption{Coupling constants for the $^{1}P_{1}~(^{3}P_{1})$ states calculated using the $^{3}P_{0}$ model. $VP$ couplings are given in units of $\mathrm{GeV}$, while $VV$ couplings are dimensionless. The values in parentheses correspond to the $^{3}P_{1}$ state.}
  \begin{ruledtabular}
    \begin{tabular}{ccc}
      Bare state& Decay channel & Coupling constant values \\\hline
      \multirow{9}{*}{$^{1}P_{1}(^{3}P_{1})$}&$D^{0}K^{*+}$&$-0.411~(0.683)$\\
      &$D^{+}K^{*0}$&$-0.340~(0.586)$\\
      &$D^{*0}K^{+}$&$-1.44~(2.09)$\\
      &$D^{*+}K^{0}$&$-1.42~(2.06)$\\
      &$D_{s}\phi$&$-2.74~(3.72)$\\
      &$D_{s}^{*}\eta$&$0.074~(-0.053)$\\
      &$D^{*+}K^{*0}$&$0.534$($/$)\\
      &$D^{*0}K^{*+}$&$0.579$($/$)\\
      &$D_{s}^{*}\phi$&$1.41$($/$)
    \end{tabular}\label{tab:qpc1}
  \end{ruledtabular}
\end{table}

\begin{figure}[!h]
  \centering
  \includegraphics[width=0.6\textwidth]{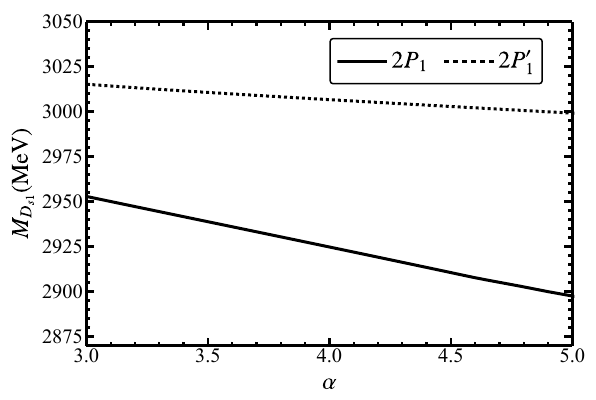}
  \caption{Variation of the physical state masses with respect to $\alpha$.}\label{fig:mass}
\end{figure}

To calculate the mass shifts, we first need the bare mass matrix. The bare mass matrix obtained from the GI model~\cite{Godfrey:2015dva} is given by
\begin{equation}
  M^{0}=\mqty(3023~\mathrm{MeV}&8.729~\mathrm{MeV}\\8.729~\mathrm{MeV}&3032~\mathrm{MeV})\,.
\end{equation}
Using the aforementioned coupling constants, we obtain $\Sigma^{T}_{ij}(m^{2},\alpha)$ and hence the effective mass matrix. The dependence of the two physical-state masses on $\alpha$ is shown in \cref{fig:mass}.
Requiring the lower eigenvalue of the effective mass matrix to equal the reference mass $m=2933~\mathrm{MeV}$, we obtain $\alpha=3.7$, which corresponds to a cutoff of $\alpha \Lambda_{\text{QCD}}=0.82~\mathrm{GeV}$. Thus, the measured mass can be accommodated with a cutoff scale in the reasonable $\sim 1~\mathrm{GeV}$ range. The corresponding mass matrix in this case is
\begin{equation}
  M=\mqty(2936~\mathrm{MeV}& 17.23~\mathrm{MeV}\\17.23~\mathrm{MeV}& 3008~\mathrm{MeV})\,.\label{eq:m1}
\end{equation}
Diagonalizing this same effective mass matrix gives a mass of $3012~\mathrm{MeV}$ for the higher-mass $2P_{1}^{\prime}$ partner state and a mixing angle of $-12.9^\circ$, which deviates from the ideal mixing limit for heavy quarks ($-35.3^\circ$).

Alternatively, keeping the same bare mass matrix, intermediate channels, $S$-wave approximation, and $^{3}P_{0}$ coupling constant $\gamma=0.4$, we examine the assignment of $D_{s1}(2933)$ as the higher-mass $2P_{1}^{\prime}$ state by requiring the higher eigenvalue of the effective mass matrix to be $2933~\mathrm{MeV}$. The other, lower-mass $2P_{1}$ state is then found at $M_{2P_{1}}=2497~\mathrm{MeV}$, with the corresponding cutoff parameter $\alpha=13.9$, or $\alpha \Lambda_{\text{QCD}}=3.1~\mathrm{GeV}$. Such a large cutoff is physically unreasonable. Moreover, in this mass region there exist $D_{s1}(2460)$ and $D_{s1}(2536)$ as $1\,{}^{1}P_{1}-1\,{}^{3}P_{1}$ partner states. It is unlikely that the $2\,{}^{1}P_{1}-2\,{}^{3}P_{1}$ multiplets can overlap with the $1\,{}^{1}P_{1}-1\,{}^{3}P_{1}$ states without impact on their properties. Therefore, we conclude that the assignment of $D_{s1}(2933)$ as the $2P_{1}^{\prime}$ state is disfavored.

The relatively small mixing angle obtained when $D_{s1}(2933)$ is identified as the lower-mass $2P_{1}$ state can be understood from the $S$-wave selection rule in the $^{3}P_{0}$ model. The $S$-wave coupling of the ${}^{3}P_{1}$ state to two vector mesons is forbidden, whereas the corresponding coupling of the ${}^{1}P_{1}$ state is allowed. Consequently, within our $S$-wave approximation, the $VV$ loops mainly contribute to the downward shift of the ${}^{1}P_{1}$ diagonal element, while their contribution to the ${}^{3}P_{1}$ diagonal element vanishes. Since the dominant $VV$ contributions do not generate a correspondingly large off-diagonal mass correction, whereas the off-diagonal correction is generated by the $VP$ channels, the off-diagonal element remains relatively small compared with the difference between the two diagonal elements. This leads to the relatively small mixing angle for the solution with the lower mass fixed at $2933~\mathrm{MeV}$.

It is worth noting that among these coupled-channel corrections, the $D^{*}K^{*}$, $D_{s}\phi$ and $D_{s}^{*}\phi$ intermediate states provide the primary contributions. This is because we only consider the $S$-wave contribution. For an $S$-wave process, there is no suppression from the momentum factor near threshold, so the channels closest to the bare-state mass dominate. Here, $D^{*}K^{*}$, $D_{s}\phi$ and $D_{s}^{*}\phi$ happen to have thresholds closest to the bare-state mass and thus provide the primary contributions.

Using the physical masses and the mixing angle, we can compute the partial widths of the primary decay channels using the $^{3}P_{0}$ model and obtain the total widths of the physical states. The decay widths are evaluated with the full $^{3}P_{0}$ amplitudes, whereas the $S$-wave restriction applies to the self-energy mass corrections. We first adopt the reference value $\gamma=0.4$ and the remaining parameters from Ref.~\cite{Godfrey:2015dva} as our baseline calculation. The results are presented in \cref{tab:width}.
\begin{table}[!t]
  \centering\caption{Partial and total widths of $D_{s1}(2P_{1}^{(\prime)})$ within the $^{3}P_{0}$ model with $\gamma=0.4$.}
  \begin{ruledtabular}
    \begin{tabular}{lcc}
      Decay channel & $\Gamma(2P_1)$ (MeV) & $\Gamma(2P_1^{\prime})$ (MeV)\\\hline
      $DK^{*}$ & 11.7 & 20.1 \\
      $D^{*}K$ & 51.2 & 40.3 \\
      $D^{*}K^{*}$ & 46.5 & 27.2 \\
      $D_{0}(1^{3}P_{0})K$ & 0.71 & 0.07 \\
      $D_{1}(1P_1)K$ & --- & 3.37 \\
      $D_{1}(1P_1^{\prime})K$ & --- & 0.89 \\
      $D_{2}(1^{3}P_{2})K$ & --- & 32.21 \\
      $D_{s}\phi$ & --- & 9.25 \\
      $D_{s}^{*}\eta$ & 4.82 & 1.39 \\
      \hline
      Total & 115& 135 \\
    \end{tabular}\label{tab:width}
  \end{ruledtabular}
\end{table}
The experimentally measured width is~\cite{LHCb:2026sup}
\begin{equation}
  \Gamma_{\text{exp}}=72^{+18}_{-12}~(\text{stat})^{+7}_{-10}~(\text{syst})~\mathrm{MeV}\,.
\end{equation}
Considering the uncertainties inherent in quark-model calculations, the predicted total width remains reasonable in comparison with the experimental measurement.

Given the sizable uncertainty in the $^{3}P_{0}$-model coupling constant, we perform a sensitivity check by determining $\gamma$ from the experimental total width rather than adopting the reference value. By requiring the calculated total width of $D_{s1}(2933)$ to be $72~\mathrm{MeV}$, we obtain $\gamma=0.32$. This value is therefore constrained by the measured width rather than being an independent prediction. The corresponding partial decay widths and total widths are listed in \cref{tab:width2}. For this value of $\gamma$, the resulting mass matrix is
\begin{equation}
  M=\mqty(2936~\mathrm{MeV}&17.17~\mathrm{MeV}\\17.17~\mathrm{MeV}&3008~\mathrm{MeV})\,,
\end{equation}
which is nearly identical to that in \cref{eq:m1} and leads to the same mixing angle, $-12.9^\circ$. The corresponding cutoff parameter is $\alpha=5.2$, which gives $\alpha \Lambda_{\text{QCD}}=1.14~\mathrm{GeV}$.

\begin{table}[!t]
  \centering\caption{Partial and total widths of $D_{s1}(2P_{1}^{(\prime)})$ within the $^{3}P_{0}$ model with $\gamma=0.32$.}
  \begin{ruledtabular}
    \begin{tabular}{lcc}
      Decay channel & $\Gamma(2P_1)$ (MeV) & $\Gamma(2P_1^{\prime})$ (MeV)\\\hline
      $DK^{*}$ & 7.31 & 12.5 \\
      $D^{*}K$ & 31.9 & 25.2 \\
      $D^{*}K^{*}$ & 29.1& 17.0 \\
      $D_{0}(1^{3}P_{0})K$ & 0.44 & 0.04 \\
      $D_{1}(1P_1)K$ & --- & 2.10 \\
      $D_{1}(1P_1^{\prime})K$ & --- & 0.56 \\
      $D_{2}(1^{3}P_{2})K$ & --- & 20.1 \\
      $D_{s}\phi$ & --- & 5.80 \\
      $D_{s}^{*}\eta$ & 3.01 & 0.87 \\
      \hline
      Total & 71.8& 84.1 \\
    \end{tabular}\label{tab:width2}
  \end{ruledtabular}
\end{table}

To compare with the experimentally measured relative branching ratio, we proceed to compute the three-body relative branching ratio for $D_{s1}(2933)$ decaying to the $DK\pi$ final state via intermediate $D_{2}^{*}K$ and $DK^{*}$ states. For this, we require the coupling constants between the physical $D_{s1}(2933)$ state and both $DK^{*}$ and $D_{2}^{*}K$. The on-shell approximation is employed for the $D_{2}^{*}K$ coupling. Using the $^{3}P_{0}$ model, we obtain
\begin{equation}
  g_{D_{s1}DK^{*}}^{\text{phys}}=1.01~\mathrm{GeV},\quad g_{D_{s1}D_{2}^{*}K}^{\text{phys}}=2.66.
\end{equation}
We then deduce $g_{K^{*}K\pi}$ and $g_{D_{2}^{*}D\pi}$ from experimental branching ratios. For $g_{K^{*} K\pi}$, we can directly use the partial width of $K^{*}\to K\pi$. The absolute branching fraction for $D_{2}^{*}\to D\pi$ has not been measured experimentally; however, we assume that single-pion production is the predominant decay mode~\cite{Falk:1992cx}. Consequently, its total width can be parameterized as
\begin{equation}
  \Gamma(D_{2}^{*})=\Gamma(D_{2}^{*}\to D \pi)+\Gamma(D_{2}^{*}\to D^{*} \pi)\,.
\end{equation}
Combining this with the measured relative branching ratio~\cite{ParticleDataGroup:2024cfk}
\begin{equation}
  \Gamma(D_{2}^{*}\to D\pi)/\Gamma(D_{2}^{*}\to D^{*}\pi)=1.52 \pm 0.14,
\end{equation}
and invoking isospin symmetry, we obtain $\Gamma(D_{2}^{*0}\to D^{+} \pi^{-})=19~\mathrm{MeV}$, which yields
\begin{equation}
  g_{K^{*}K\pi}=4.40,\quad g_{D_{2}^{*}D\pi}=25.7~\mathrm{GeV}^{-1}.
\end{equation}
Finally, we obtain the theoretical relative three-body branching ratio. Since the common $^{3}P_{0}$ coupling factor $\gamma$ cancels between the numerator and denominator, this ratio is independent of whether $\gamma=0.4$ or $0.32$ is used:
\begin{equation}
  R_{\text{th}}=\frac{\Gamma(D_{s1}(2933)\to D^{+}(K^{*0}\to K^{+}\pi^{-}))}{\Gamma(D_{s1}(2933)\to K^{+}(D_{2}^{*}(2460)^{0}\to D^{+}\pi^{-}))}=1.98\,,
\end{equation}
whereas the corresponding experimental value is given by~\cite{LHCb:2026sup}
\begin{equation}
  R_{\text{exp}}= \frac{\mathrm{BR}(B^{0}\to D^{-} D_{s1}(2933)\to D^{+}K^{*0}(\to K^{+}\pi^{-}))}{\mathrm{BR}(B^{0}\to D^{-} D_{s1}(2933)\to K^{+}D_{2}^{*0}(\to D^{+}\pi^{-}))}=1.27^{+1.29}_{-0.95}\,.
\end{equation}
Although our theoretical prediction is about 1.5 times the central experimental value, the theoretical prediction and experimental measurement are consistent within the large experimental uncertainties. In conclusion, our theoretical calculation for the three-body relative branching ratio provides further support for identifying $D_{s1}(2933)$ as the $2P_{1}$ state predicted by the quark model, but with rather significant influences from the coupled-channel effects.

\section{Summary}\label{sec:summary}
In this work, the newly observed $D_{s1}(2933)$ state at LHCb has been investigated as the $2P_{1}$ excited state of the $c\bar{s}$ system, which is affected by the nearby $S$-wave coupling channels. With the bare state from the Godfrey-Isgur quark model, its couplings to the nearby $S$-wave channels are calculated in the $^{3}P_{0}$ model.  By including the coupled-channel corrections induced by the $DK^{*}$, $D^{*}K$, $D^{*}K^{*}$, $D_{s}\phi$, $D_{s}^{*}\phi$ and $D_{s}^{*}\eta$ intermediate loops, the bare masses of the $2P_{1}^{(\prime)}$ states from the GI model can be shifted down to the measured value. By assigning the eigenvalue of the lower state to the measured mass, it gives the higher partner state $2P_{1}^{\prime}$ at $3012~\mathrm{MeV}$ and a $^{1}P_{1}$--$^{3}P_{1}$ mixing angle of $-12.9^{\circ}$. The masses of these two $2P$ states deviate significantly from the masses predicted by the GI model, and the extracted mixing angle is also different from the ideal-mixing angle of $-35.3^{\circ}$ based on the heavy-quark symmetry.  In contrast, assigning $D_{s1}(2933)$ to the higher-mass $2P_{1}^{\prime}$ state requires a much larger cutoff scale of $3.1~\mathrm{GeV}$ and is therefore disfavored. Since only the $S$-wave contributions are considered in the present analysis, the channels with thresholds closest to the bare-state mass provide the dominant contributions to the mass shift.

We also investigate the total and partial decay widths in the quark model. It shows that with the quark model coupling $\gamma=0.32$, a total width of $71.8~\mathrm{MeV}$ can be obtained, which is in agreement with the experimental data $\Gamma_{\mathrm{exp}}=72^{+18}_{-12}(\mathrm{stat})^{+7}_{-10}(\mathrm{syst})~\mathrm{MeV}$. The value of $\gamma=0.32$ is within the range of uncertainties of the quark model parameters. Note that $\gamma$ is an overall parameter, it should be more conclusive to look at the branching ratio fractions between exclusive decay channels.
We find that the predicted three-body branching ratio fraction for $D_{s1}(2933)\to D^{+}K^{+}\pi^{-}$ through the $DK^{*}$ and $D_{2}^{*}K$ intermediate states, $R_{\mathrm{th}}=1.98$, is consistent with the experimental value $R_{\mathrm{exp}}=1.27^{+1.29}_{-0.95}$ within the sizable uncertainties. Together with the total width, these results suggest the preference of  $D_{s1}(2933)$ as the lower $2P_{1}$ state of the $c\bar{s}$ system, but strongly affected by the nearby $S$-wave coupling channels.
More precise measurements of the partial branching fractions at LHCb and BESIII, particularly the ratio $\Gamma(D_{s1}(2933)\to D^{*}K)/\Gamma(D_{s1}(2933)\to DK^{*})$, would provide a more stringent test of this assignment and facilitate the search for the $2P_{1}^{\prime}$ partner state.

\acknowledgments
This work is supported by the National Natural Science Foundation of China (Grant No. 12235018).

\bibliography{ref-Ds1(2933).bib}
\end{document}